\documentclass[proceedings]{JHEP3}

\PrHEP{PrHEP hep2001}		
\conference{International Europhysics Conference on HEP}

\usepackage{epsfig} 

\newcommand{\beq}{\begin{equation}} 
\newcommand{\eeq}{\end{equation}} 
\newcommand{\beqa}{\begin{eqnarray}} 
\newcommand{\eeqa}{\end{eqnarray}} 
\newcommand{\beqan}{\begin{eqnarray*}} 
\newcommand{\eeqan}{\end{eqnarray*}} 
\newcommand{\ba}{\begin{array}} 
\newcommand{\ea}{\end{array}}

\newcommand{\bea}{\begin{eqnarray}} 
\newcommand{\eea}{\end{eqnarray}}

\newcommand{\PL}[3]{{Phys. Lett.} {\bf#1} {(#2)} {#3}} 
\newcommand{\PRL}[3]{{Phys. Rev. Lett.}  {\bf#1} {(#2)} {#3}} 
\newcommand{\PR}[3]{{Phys. Rev.} {\bf#1} {(#2)} {#3}} 
\newcommand{\NP}[3]{{Nucl. Phys.} {\bf#1} {(#2)} {#3}} 
\newcommand{\EPJ}[3]{{Eur. Phys. J.} {\bf#1} {(#2)} {#3}}

\title{Isospin Violation and the Magnetic Moment of the Muon}

\author{\speaker{Vincenzo Cirigliano}, Gerhard Ecker, Helmut Neufeld 
        \thanks{ Work supported in part by TMR, EC-Contract  
        No. ERBFMRX-CT980169 (EURODA$\Phi$NE).}\\  
	Institut f\"ur Theoretische Physik, Universit\"at 
        Wien\\ Boltzmanngasse 5, A-1090 Vienna, Austria \\
        E-mail: \email{vincenzo@thp.univie.ac.at}}	

\abstract{We calculate the leading isospin-violating and electromagnetic
corrections for the decay $\tau^- \to \pi^0 \pi^- \nu_\tau$ at low
energies. The corrections are small but
relevant for the inclusion of $\tau$ decay data in the determination 
of hadronic vacuum polarization especially for the anomalous magnetic
moment of the muon. This contribution is based on Ref.~\cite{cen}}

\begin{document}

\section{Introduction}
The leading hadronic contribution to the anomalous magnetic moment of
the muon $a_\mu=(g_\mu-2)/2$ is given by the hadronic vacuum
polarization \cite{GdR69}:
\begin{equation}
a_\mu^{\rm vacpol}=\displaystyle\frac{1}{4\pi^3}\displaystyle\int_{4
M_\pi^2}^{\infty} dt K(t) \sigma^0_{e^+ e^- \to  {\rm hadrons}}(t)
\end{equation} 
where $K(t)$ is a smooth kernel concentrated at low energies,
and $\sigma^0_{e^+ e^- \to  {\rm hadrons}}$ denotes the 
``pure'' hadronic cross section with QED corrections removed.
The low-energy structure of hadronic vacuum polarization is especially
important. In fact, about 70 $\%$ of $a_\mu^{\rm vacpol}$ is due to
the two-pion intermediate state for $4 M_\pi^2 \le t \le 0.8$ GeV$^2$
(see, e.g., Ref. \cite{narison}).  A precision of 1 $\%$ has been
achieved in the calculation of $a_\mu^{\rm vacpol}$ by including
\cite{alemany} the more accurate $\tau$ decay data \cite{taudata} in
addition to $\sigma(e^+ e^- \to $ hadrons) data. 
This is possible because of a CVC relation between
electromagnetic and weak form factors in the isospin limit. However,
both the aforementioned theoretical accuracy and the new
high-precision experiment at Brookhaven \cite{bnl} warrant a closer
investigation of isospin violation, due to both the light quark
mass difference and electromagnetism (EM).
We concentrate in this work on isospin violation in the reactions
$\tau^- \to \pi^0 \pi^- \nu_\tau$ and $e^+ e^- \to \pi^+ \pi^-$ at low
energies. Chiral perturbation theory (CHPT) \cite{chpt,urech} is the
framework where such corrections can reliably be calculated for the
standard model in a systematic low-energy expansion. More
specifically, we calculate the leading corrections of both
$O[(m_u-m_d)p^2]$ and $O(e^2 p^2)$ for the CVC relation between the
two-pion (vector) form factors in the two processes. 
For a more detailed presentation and a more complete bibliography 
we refer to Ref.~\cite{cen}. 

Let us define the problem more precisely.  For the two-pion final
state, the bare $e^+ e^-$ cross section and the $\tau^- \to \pi^0 \pi^-
\nu_\tau$ decay distribution take the form:
\begin{eqnarray} 
\sigma^0_{e^+ e^- \to \pi^+ \pi^-}(t)&=&\displaystyle\frac
{\pi\alpha^2}{3 t}  \beta^3_{\pi^+\pi^-} (t) |F_V(t)|^2 \\ 
\displaystyle\frac{d \Gamma(\tau^-\to \pi^0 \pi^- \nu_\tau)}{dt} &=& 
\Gamma_e \, {\cal K}_1 (t) \, \beta^3_{\pi^0\pi^-}(t) \, 
|f_+(t)|^2 \, S_{\rm EW} \, G_{\rm EM} (t) \ , 
\label{dGamma} 
\end{eqnarray} 
where 
$$
\Gamma_e= \displaystyle\frac{G_F^2 m_\tau^5}{192 \pi^3} ~,~ \ \ \ \ \ \
{\cal K}_1(t) =  
\displaystyle\frac{|V_{ud}|^2}{2 m_\tau^2} (1-\frac{t}{m_\tau^2})^2  
(1+\frac{2t}{m_\tau^2}) \ . 
$$
Here $t$ is the hadronic invariant mass; $\beta_{\pi^+ \pi^-} (t)$ and
$\beta_{\pi^0 \pi^-} (t)$ are the center of mass pion velocities for
the two processes; $F_V (t)$ and $f_+ (t)$ are the EM and weak vector
form factors of the pion. In the isospin limit ($m_u=m_d$ and $e=0$)
we have $M_{\pi^+}=M_{\pi^0}$ (hence $\beta_{\pi^+ \pi^-} (t)=
\beta_{\pi^0 \pi^-} (t)$) , $S_{\rm EW} = G_{\rm EM} (t) = 1$ and
$f_+(t) = F_V(t)$, implying the CVC relation
\begin{equation} 
\sigma^{0, CVC}_{e^+ e^- \to \pi^+ \pi^-}(t)=\displaystyle\frac{\pi \alpha^2}
{3 \, t \,  {\cal K}_1(t) \  \Gamma_e} 
\displaystyle\frac{d \Gamma(\tau^-\to \pi^0 \pi^-
\nu_\tau)}{dt} \ . 
\label{CVC}
\end{equation} 
Including isospin violation, the modified CVC relation reads
\begin{eqnarray} 
\sigma^0_{e^+ e^- \to \pi^+ \pi^-}(t) & = & 
\sigma^{0, CVC}_{e^+ e^- \to \pi^+ \pi^-}(t) 
\displaystyle\frac{R_{\rm IB} (t)}{S_{\rm EW}} \ ,  \\
 R_{\rm IB} (t) & = &   
\displaystyle\frac{1}{G_{\rm EM}(t)} \displaystyle\frac{\beta^3_{\pi^+ 
\pi^-}(t)}{\beta^3_{\pi^0\pi^-}(t)} \left|\displaystyle\frac{F_V(t)}{f_+(t)}
\right|^2 ~. \label{riso}
\end{eqnarray}
The factor $S_{\rm EW}$ takes into account the dominant short-distance
electroweak corrections \cite{MS88}.  To lowest order in $\alpha$, it
is given by $S_{\rm EW} = 1 +(\alpha / \pi) {\rm
log}(M_Z^2/m_{\tau}^2)$, amounting to the numerical value $S_{\rm EW}
= 1.0194$. This is consistently used in all present analyses. 
$R_{\rm IB} (t)$  involves the long distance QED factor $G_{\rm EM}(t)$ (which
receives both virtual and real photon contributions), the phase space
correction factor $\beta^3_{\pi^+ \pi^-}(t)/\beta^3_{\pi^0\pi^-}(t)$
~\cite{kuehn}, and the ratio of EM and weak form factors.  Working at
leading order, the form factor $F_V(t)$ needs to be calculated to
$O[(m_u-m_d)p^2]$ with physical meson masses (but without explicit
photonic corrections). $f_+(t)$ must be calculated to both
$O[(m_u-m_d)p^2]$ and $O(e^2 p^2)$, if $d\Gamma/dt$ is to be extracted
from actual $\tau$ decay data. 

\section{Anatomy of $R_{\rm IB}(t)$: $F_V$, $f_+$, $G_{\rm EM}$}

At first non-trivial order in the low-energy expansion, isospin
violation manifests itself in the pion vector form factor $F_V(t)$
only in the masses of the particles contained in the loop amplitude:
\begin{equation} 
F_V(t)=1+2 H_{\pi^+\pi^-}(t) + H_{K^+ K^-}(t)
\label{FVp4}
\end{equation}
with \cite{gl852}
\begin{equation} 
H_{PQ} (t) =  {\tilde H}_{PQ}(t,\mu) 
+ \frac{2}{3 F_\pi^2}t L_9^r(\mu) ~, 
\end{equation} 
The loop function ${\tilde H}_{PQ}(t)$ encodes the singularities due
to the low energy meson propagation, while the local term
(proportional to the low-energy constant $L_9^r(\mu)$ \cite{gl852})
governs the charge radius of the pion and is sensitive to the
structure of the theory at higher energies. 
This specific channel is completely
dominated by the $\rho$ resonance.  So, leaving the domain of a pure
low energy effective theory, we use the prescription of
Ref.~\cite{gpgdpp} to match the CHPT form factor (\ref{FVp4}) of
$O(p^4)$ to the resonance region:
\begin{eqnarray} 
F_V(t)&=&\displaystyle\frac{M_\rho^2}{M_\rho^2 - t -i M_\rho
\Gamma_\rho(t)}  \exp{ \bigg[2 {\rm Re} {\tilde H}_{\pi^+\pi^-}(t)+
{\rm Re} {\tilde H}_{K^+ K^-}(t) \bigg]}~,
\label{FVrho} \\
\Gamma_\rho(t)&=&\displaystyle\frac{M_\rho t}{96 \pi F_\pi^2}
\left[\beta^3_{\pi\pi}(t)\theta(t-4 M_\pi^2)+\frac{1}{2}
\beta^3_{KK}(t) \theta(t-4 M_K^2)\right] \ . 
\label{width}
\end{eqnarray}
For the present case, the charged pion and kaon masses must be inserted 
in the hadronic off-shell width. 
The representation (\ref{FVrho}) gives an excellent description of
$e^+ e^- \to \pi^+ \pi^-$ data up to $t\sim 1$ GeV$^2$ with the single
parameter $M_\rho\simeq$ 775 MeV, and has the correct low-energy
behaviour to $O(p^4)$ (including isospin breaking).
To the order we are working, $\rho^+-\rho^0$ mass difference and
$\rho-\omega$ mixing do not appear. Such higher-order effects (from
the low energy point of view) are not necessarily negligible
numerically (see, e.g., Ref.~\cite{maltman}).  They can be included 
as additional contributions to the factor $R_{\rm IB}(t)$. 

To first order in isospin violation, this time including explicit
photonic corrections, the form factor $f_+$ is given by
\begin{equation} 
f_+(t,u)=1 + 2 H_{\pi^0\pi^-}(t) + H_{K^0 K^-}(t)
           +  f_{\rm loop}^{\rm elm}(u,M_\gamma)
+ f_{\rm local}^{\rm elm}~. 
\label{fplus1}
\end{equation}
Compared to the form factor $F_V(t)$ in (\ref{FVp4}), the appropriate
meson masses appear in the loop amplitude and there is an additional
electromagnetic amplitude, containing both the photon loop diagrams
and an associated local part. The electromagnetic amplitude depends on
the second Dalitz variable $u=(P_\tau - p_{\pi^-})^2$.
The loop function $f_{\rm loop}^{\rm elm}(u,M_\gamma)$ encodes
universal physics related to the Coulomb interaction between the
$\tau$ and the charged pion, and therefore we pull it out in an overall 
term. Matching this low energy result to the resonance region, 
 we are lead to write:
\begin{eqnarray} 
f_+(t,u) & = & f_{+} (t) \bigg[1 + f_{\rm loop}^{\rm elm}(u,M_\gamma)
\bigg] \\ 
f_+(t)&=&\displaystyle\frac{M_\rho^2}{M_\rho^2 - t -i M_\rho
\Gamma_\rho(t)}  \exp{ \bigg[2 {\rm Re} {\tilde H}_{\pi^0\pi^-}(t)+
{\rm Re}{\tilde H}_{K^0 K^-}(t) \bigg]} + f_{\rm local}^{\rm elm} ~.
\label{fplus2}
\end{eqnarray}
The resonance width $\Gamma_\rho (t)$ in (\ref{width}) has to
be calculated now with the appropriate $\pi^- \pi^0$ and $K^- K^0$
thresholds and phase space factors. 
The local contribution in $f_+$ depends on three low-energy constants
appearing in the chiral expansion. The bounds used for them 
\cite{cen} reflect in
the uncertainty reported in our result (Fig.~\ref{fig:RIB} $(a)$,
solid curves).

The photon loop amplitude $f_{\rm loop}^{\rm elm}(u,M_\gamma)$ is
infrared divergent, depending on an artificial photon mass
$M_\gamma$. This dependence is canceled by bremsstrahlung of soft
photons making the decay distribution in $(t,u)$ infrared finite.  The
sum of real and virtual contributions produces a correction factor
$\Delta (t,u)$ to the $(t,u)$ decay distribution.  After averaging
over the Dalitz variable $u$, this produces the term $G_{\rm EM}(t)$
in the decay distribution (\ref{dGamma}).  The precise form of
$\Delta (t,u)$ (and $G_{\rm EM}(t)$) depends on the specific
experimental setup.  To the best of our knowledge, all $\tau$ decay
experiments relevant here \cite{taudata} apply bremsstrahlung
corrections in the same (approximate) way ~\cite{was}, including only
the leading term in the Low expansion. Assuming this prescription, we
have then calculated the setup-independent part of $G_{\rm EM}(t)$,
assigning to it an uncertainty of $\pm \alpha/\pi$ (due to neglect of
sub-leading terms).

\section{Results and conclusions}
\begin{figure}
\centering
\begin{picture}(300,180)  
\put(-5,50){\makebox(100,120){\epsfig{figure=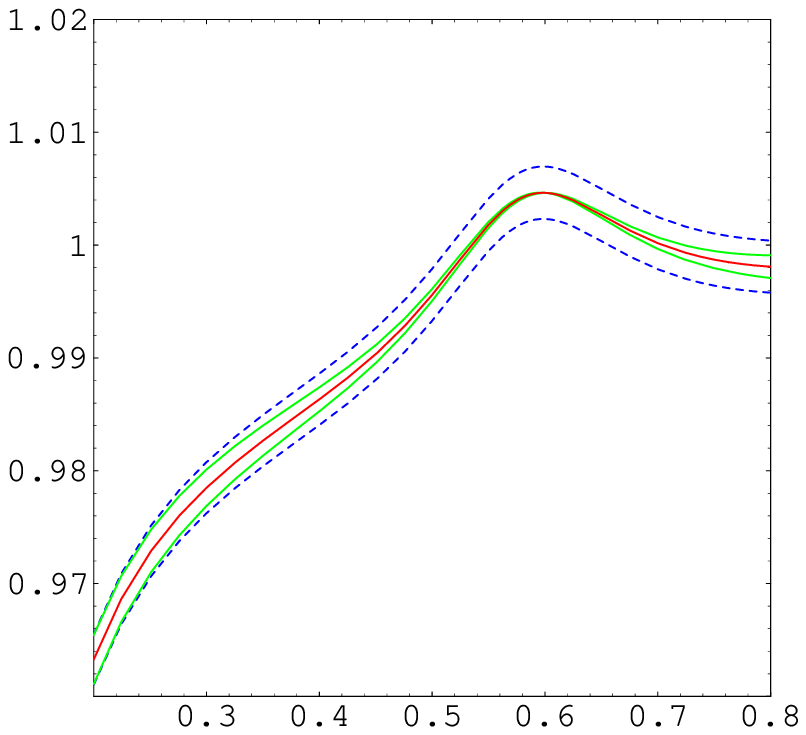,height=6.0cm}}}
\put(120,20){\scriptsize{$t$ (GeV$^2$)}}
\put(-65,170){\scriptsize{$R_{\rm IB} (t)$}}
\put(40,20){\scriptsize{$(a)$}}
\put(220,50){\makebox(100,120){\epsfig{figure=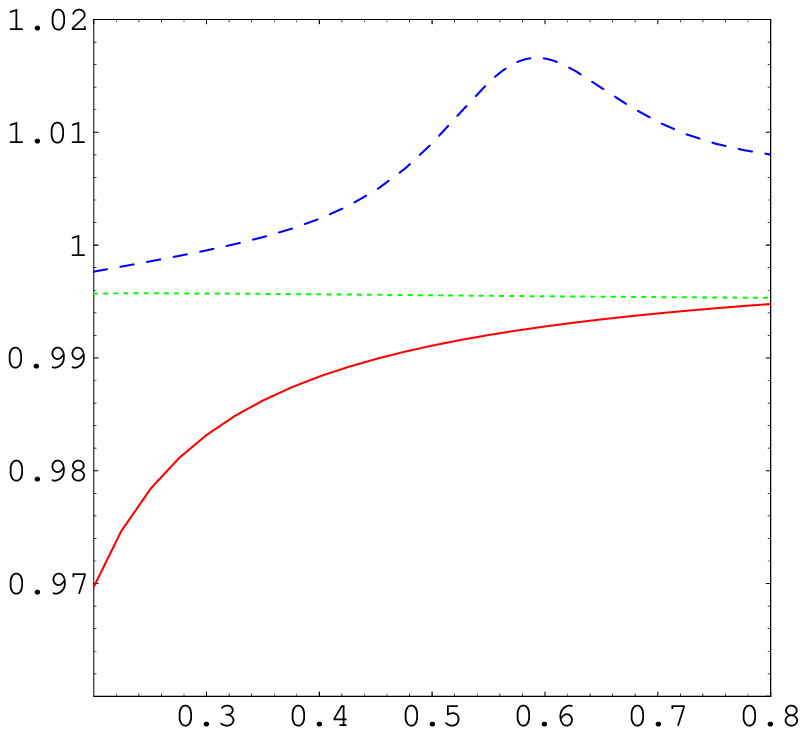,height=6.0cm}}}
\put(320,20){\scriptsize{$t$ (GeV$^2$)}}
\put(280,20){\scriptsize{$(b)$}}
\end{picture}
\caption{$(a)$ Correction factor $R_{\rm IB}(t)$ for isospin violation. 
The bands around the central curve correspond to
the uncertainty in the low-energy constants (solid lines)  
and in the bremsstrahlung factor (dashed lines).  
$(b)$ The separate factors defining  $R_{\rm IB}(t)$ in 
Eq.(\protect\ref{riso}) are plotted as solid line for
$\beta_{+-}^3/\beta_{0-}^3$, dashed line for $|F_V(t)/f_+(t)|^2$ and
dotted line for $1/G_{\rm EM}(t)$. \label{fig:RIB} }
\end{figure}
The results of our analysis are summarized in Figs.~\ref{fig:RIB}
$(a)$,$(b)$ where we plot the function $R_{\rm IB}(t)$ of 
Eq.~(\ref{riso}) and its component factors for $ 0.2 \leq t \leq 0.8 $
GeV$^2$.  We note that the dominant contribution at low $t$ is given
by the kinematical term $\beta_{+-}^3/\beta_{0-}^3$
\cite{kuehn}. Photonic corrections embodied in $G_{\rm EM}(t)$ reduce
$R_{\rm IB}(t)$ in addition by about half a percent, largely
independently of $t$. The form factor ratio $|F_V(t)/f_+(t)|^2$ is
dominated by the width difference $\Gamma_{\rho^+}-\Gamma_{\rho^0}$.   
Since $R_{\rm IB}(t)$ is smaller than unity in most of the region
under consideration ($4 M_\pi^2 \le t \le 0.8$ GeV$^2$), isospin
violation accounts for a sizable part of the systematic difference at
low energies between $e^+ e^-$ and $\tau$ decay data (e.g.,
Ref.~\cite{EI99}). \\
In order to quantify the impact of $R_{\rm IB}(t)$ on 
$a_\mu^{\rm vacpol}$, we construct the following ratio: 
\begin{equation}
{\cal R} (t_{\rm max}) = \frac{ \displaystyle\int_{4 M_\pi^2}^{t_{\rm
max}} dt \, K(t) \, \sigma^{0,{\rm CVC}}_{e^+ e^- \to \pi^+ \pi^-  }(t) \, 
R_{\rm IB}
(t)}{ \displaystyle\int_{4 M_\pi^2}^{t_{\rm max}} dt \, K(t) \,
\sigma^{0,{\rm CVC}}_{e^+ e^- \to \pi^+ \pi^- }(t) } ~ , 
\label{ramu}
\end{equation}
and report a few representative values of ${\cal R} (t_{\rm max})$ in 
Table \ref{tab1}. 
\renewcommand{\arraystretch}{1.5}
\TABLE{
\caption{Correction factor for $a_\mu^{\rm vacpol}$ due to isospin 
violation (defined in 
Eq.~(\protect\ref{ramu})) for some values of $t_{\rm max}$.
An uncertainty of $0.002$ - due to  $G_{\rm EM}(t)$ - 
should be assigned to the reported values.
This is also an upper bound for the uncertainty due to the low-energy 
constants (see Fig. \ref{fig:RIB}(a)). \label{tab1} } 
\begin{tabular}{|c||c|c|c|}\hline
$t_{\rm max}$ (GeV$^2$) & 0.3 & 0.5 & 0.8\\ \hline
${\cal R} (t_{\rm max})$ & 0.949 & 0.974 & 0.988 \\  
\hline 
\end{tabular}
}
%
Although the calculation is based on a low-energy 
description of the standard model, we claim that the main features of
our $R_{\rm IB}(t)$ are valid up to $t \simeq 0.8$
GeV$^2$. Of the three factors in the definition (\ref{riso}) of
$R_{\rm IB}(t)$, both the dominant phase space correction factor
\cite{kuehn} and the photon loop effects are independent of the
low-energy expansion. Finally, the main part of isospin violation in
the form factor ratio $|F_V(t)/f_+(t)|^2$ occurs in the $\rho$-width
difference $\Gamma_{\rho^+}-\Gamma_{\rho^0}$ and should therefore be
reliable in the vicinity of the resonance.



\end{document}